# In-Situ Monitoring of Thermal Annealing Induced Evolution in Film Morphology and Film-Substrate Bonding in a Monolayer MoS$_2$ Film

Liqin Su[1], Yifei Yu[2], Linyou Cao[2], and Yong Zhang[1,*]

[1]*Department of Electrical and Computer Engineering, University of North Carolina at Charlotte, Charlotte, NC 28223*

[2]*Department of Materials Science and Engineering, North Carolina State University, Raleigh, NC 27695*

\* To whom correspondence should be addressed: yong.zhang@uncc.edu

**Abstract**

We perform in-situ two-cycle thermal cycling and annealing studies for a transferred CVD-grown monolayer $MoS_2$ on a $SiO_2$/Si substrate, using spatially resolved micro-Raman and PL spectroscopy. After the thermal cycling and being annealed at 305 °C twice, the film morphology and film-substrate bonding are significantly modified, which together with the removal of polymer residues cause major changes in the strain and doping distribution over the film, and thus the optical properties. Before annealing, the strain associated with ripples in the transferred film dominates the spatial distributions of the PL peak position and intensity over the film; after annealing, the variation in film-substrate bonding, affecting both strain and doping, becomes the leading factor. This work reveals that the film-substrate bonding, and thus the strain and doping, is unstable under thermal stress, which is important for understanding the substrate effects on the optical and transport properties of the 2D material and their impact on device applications.

# Introduction

Single layer Molybdenum disulfide ($MoS_2$) along with other two-dimensional (2D) materials has been shown exhibiting many unique electronic and optical properties. However, their properties are very sensitive to the perturbations from either supporting substrates or surface contaminants as well as unintended film morphology fluctuations. Among these external perturbations, the role of the substrate is "intrinsic" and thus ultimately important to fundamental understanding and application of the 2D material. Numerous studies on the substrate supported transition metal dichalcogenide (TMD) films concerned the effects of different substrate types.[1-5] It was often implicitly assumed that the substrate effects depended solely on the substrate material and the film-substrate bonding was stationary, thus, the obtained results were representative of the specific substrate type. However, it has been shown that on the same substrate type variations in film morphology and film-substrate bonding strength can have major impacts on the material properties.[2, 4] Furthermore, the film-substrate interaction have been shown to be non-stationary under thermal[2, 4] and electrical stresses,[6-8] and may differ significantly for the same substrate material but prepared differently (e.g., epitaxially grown vs. transferred).[4] Changes in surface morphology and film-substrate bonding during thermal annealing have been attributed to the unusual temperature evolution in the optical properties,[2, 4] whereas alternations in the interfacial states and surface contaminants under electrical stress have been suggested to be responsible for the instability of electrical characteristics.[6-8] Therefore, the answer to a question like how the substrate will impact the carrier saturation velocity of a 2D material is unlikely to be unique.[9-11] Furthermore, it is unclear how thermal annealing or unintended thermal stress during temperature dependent measurements will affect the material properties, and how will be the interplays of above mentioned "intrinsic" (substrate) and extrinsic perturbations responding to the annealing.

We have previously studied the substrate effects in one single low-to-high temperature cycle up to the thermal degradation point (typically > 500 °C) for $MoS_2$ and $WS_2$, which has provided valuable information about the film-substrate interaction as mentioned above.[2, 4] However, these efforts could not provide the material properties at room temperature (RT) after the thermal cycle. In this work, we perform an in-situ two-cycle thermal annealing study, with the upper temperature limited to 305 °C (much below the degradation point), on both Raman and PL characteristics of a transferred monolayer $MoS_2$ film on a $SiO_2$/Si substrate. Specifically, the measurements are carried out at RT before and after the first cycle, and after the second cycle, and during the first and second cycle. This effort allows us to reveal the annealing effects on the film morphology, film-substrate bonding, and surface contaminants, the consequences of annealing to strain and doping, and the manifestations on the optical properties. Additionally, using spatially resolved μ-Raman and PL we are able to study the spatial inhomogeneity of these effects. The findings have major implication on the understanding of the electronic transport properties, and tuning the material properties through substrate engineering.

The influence of the substrate on the electrical and optical properties of 2D films is typically associated with the strain and doping effects.[12-14] Raman spectroscopy is often used to probe these two effects in the TMD films, because the two primary Raman modes, in-plane $E_{2g}$ mode and out-of-plane $A_{1g}$ mode, respond differently to the two effects: $E_{2g}$ is more sensitive to the strain than $A_{1g}$,[15] while $A_{1g}$ is much more sensitive to the doping than $E_{2g}$.[16] The deformation potentials under biaxial strain for the phonon modes and bandgap have been estimated to be −4.5 cm$^{-1}$/% for $E_{2g}$, −1.0 cm$^{-1}$/% for $A_{1g}$,[17] and −70 meV/% for bandgap,[18] respectively. Electron doping results in red-shifts of both Raman modes, but the $E_{2g}$ shift is about 1/9 of $A_{1g}$.[16] Structural defects may introduce bound states that could either provide doping in the 2D film if they are shallow, or quench the interband recombination when they are far away from the band edges.[19] It has been proposed that charge transfer between the film and the substrate can significantly modify the doping concentration, and influence the optical properties.[14, 20, 21] These excessive charged carriers may couple with neutral excitons to form trions: $A^-$ (negatively charged) and $A^+$ (positively charged).[22] Furthermore, the monolayer $MoS_2$ is often grown by CVD and then transferred to another substrate with polymer-assisted transfer processes, leaving behind residuals on the surface of $MoS_2$ which is challenging to be removed. The polymer residues as well as adsorbed $H_2O$ and $O_2$ are known to modify the optical and electrical properties of the film, such as the quenching of photo-generated excitons and the reduction of carrier mobility.[5, 23-25]

Temperature dependent Raman scattering has been used to investigate the vibrational properties of both bulk and monolayer $MoS_2$, and in general both $E_{2g}$ and $A_{1g}$ peaks exhibit red-shift with increasing temperature.[2, 26-28] Besides the fundamental interest in the vibrational properties of the 2D material, we have shown that because of the expected nearly linear temperature dependence of the phonon frequencies for an idealistic 2D material in the elevated temperature region (above RT), the temperature dependent Raman study can be used as an effective probe to the film-substrate interaction.[2, 4, 29] For instance, in our previous work we have shown that the $A_{1g}$ mode, as well as the $E_{2g}$ mode to a less extent, shows an anomalous nonlinear temperature dependence due to temperature-induced changes in the film morphology and the interaction with the substrate.[2] The similar effect occurred in graphene, which limited the study of the intrinsic temperature dependence in a lower temperature region.[30]

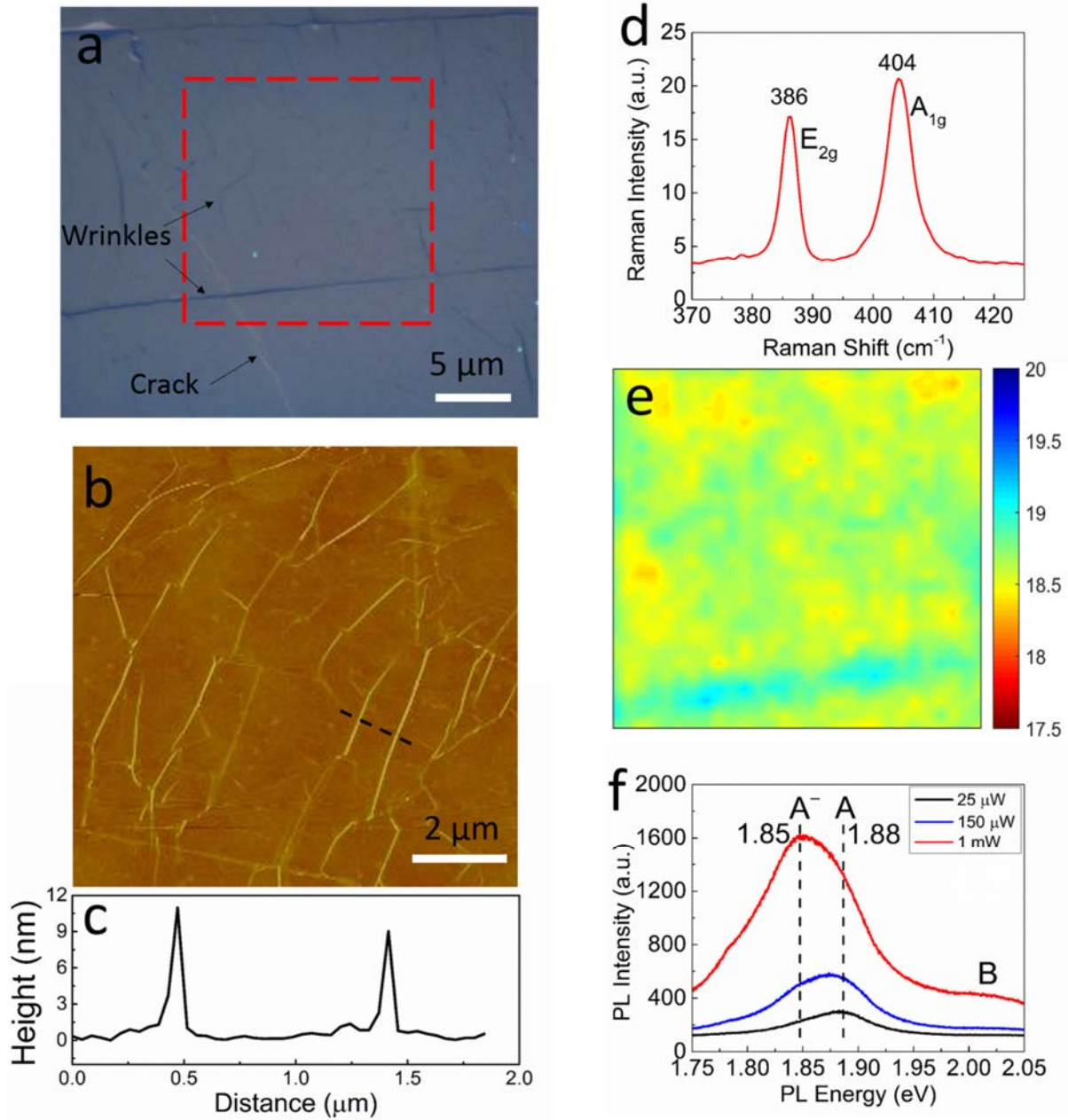

**Figure 1.** Room temperature characterization of a transferred monolayer MoS$_2$ film on a SiO$_2$/Si substrate. (a) Optical image. (b) Atomic force microscopy (AFM) image (not the same area as the marked one). (c) The height profile extracted from AFM image along the dash line labelled in panel b. (d) Typical Raman spectrum. (e) The spatial mapping (20 μm × 20 μm) of the frequency difference between E$_{2g}$ and A$_{1g}$ in the area labelled in panel a. (f) Typical power dependent PL spectra.

**Results and Discussion**

Monolayer MoS$_2$ film was originally grown on sapphire by CVD, and then transferred onto a Si substrate coated with 300 nm thick SiO$_2$ with the assistance of polystyrene (PS).[31] The as-transferred film exhibits weak bonding with the substrate.[5] Figure 1a shows the optical image of a transferred MoS$_2$ film on which spatially resolved Raman and PL measurements were performed in a marked area of 20 μm × 20 μm. The sample contains a visible long wrinkle (in darker color near the bottom of the marked area) and many less visible shorter ones, and a crack (in lighter color near the lower left corner of the marked area). In order to see the shorter wrinkles more apparently, an AFM image (not the same area as the marked one) is shown in Figure 1b, and the wrinkles are less than 12 nm in height (Figure 1c). The primary focus of this work is to understand the behavior of the general area, while the behavior of the wrinkle will only be briefly addressed when appropriate. Raman spectroscopy is a convenient and effective method to determine the thickness of the film through the frequency difference between A$_{1g}$ and E$_{2g}$ modes – the difference should be < 20 cm$^{-1}$ for monolayer.[32] Figure 1d shows a typical Raman spectrum of the sample, and the frequencies of E$_{2g}$ and A$_{1g}$ are ~386 cm$^{-1}$ and ~404 cm$^{-1}$, respectively. The spatial mapping of the frequency difference over the marked area in Figure 1a shows a maximum of 19.2 cm$^{-1}$ (Figure 1e), indicating the film overall is indeed monolayer. Figure 1f shows RT PL spectra at different laser powers. At 25 μW only one peak at ~1.88 eV is observed; however, a new peak appears at ~1.85 eV as the power increases to 150 μW, and then dominates the PL spectrum as the power increases further to 1 mW. The lower-energy component cannot be assigned as the emission related to impurity or defect states. If it were the case, at the low excitation level, the electrons would tend to occupy these states prior to the conduction band, thus the lower energy peak would be dominant; furthermore, the intensity ratio of the lower-energy peak to the higher-energy peak would decrease with increasing laser power, because of the state filling effect. On the contrary, the proportion of the lower-energy peak increases with increasing laser power. As mentioned above, the free carriers can couple with neutral excitons to form trions, and the emission of trions increases with increasing carrier density.[14, 33] Therefore, the lower-energy peak at 1.85 eV can be assigned as $A^-$. Another weak PL peak at ~2.02 eV can also be seen, and is assigned as B exciton. The A and B excitons originate from the splitting of the valence band at the K point.[34]

Figure 2 presents the RT Raman and PL mapping results of integrated intensity and peak position before annealing the film. The Raman intensity data (Figure 2a-b) show an overall uniform distribution over the film except for along the wrinkles with higher intensity and near the crack with lower intensity. The wrinkled regions (appearing as lines in optical and AFM images) tend to have larger effective absorbing areas, resulting in stronger Raman intensity. The PL intensity mapping shows a more significant variation, ~20% (Figure 2c). The Raman and PL peak position data are shown in Figure 2d-f, with maximum variations of ~0.6 cm$^{-1}$ for $E_{2g}$ mode, ~0.7 cm$^{-1}$ for $A_{1g}$ mode, and ~12 meV for PL, respectively. There appears a general correlation, revealed by the similar patterns, between $E_{2g}$ Raman frequency and PL energy over the mapped area, i.e. the area with lower (higher) Raman frequency (Figure 2d) shows lower (higher) PL energy (Figure 2f). Interestingly, the wrinkles shown in the marked area (Figure 1a) generally match the areas with lower $E_{2g}$ frequency and PL energy. Therefore, the origin of the non-uniform $E_{2g}$ frequency and PL energy is most likely due to the morphology fluctuations over the MoS$_2$ monolayer, i.e. the transferred film was not laid down perfectly flat but with many microscopic scale ripples and elongated wrinkles, where wrinkles can be considered as ripples with more abrupt changes in morphology. Thus, if the strain is responsible for the variations, using the deformation potentials given above, the corresponding maximum strain differences derived from the data of Figure 2d-f,

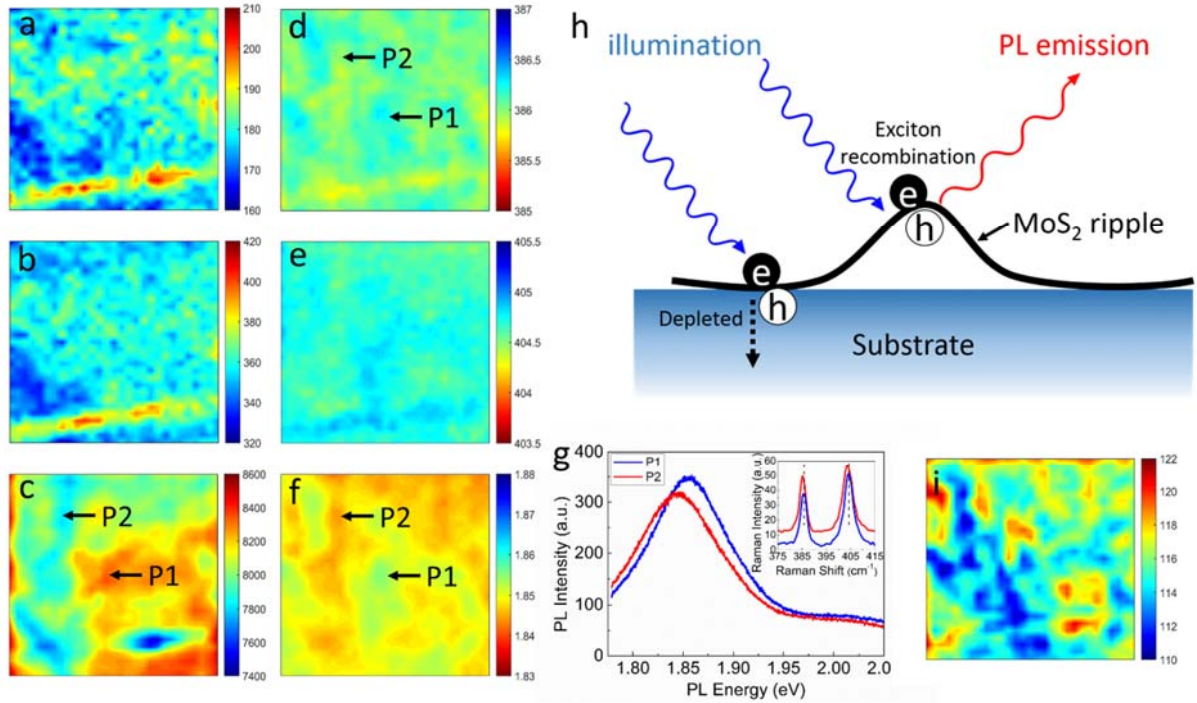

**Figure 2.** Room temperature Raman and PL mapping results before the first thermal annealing. (a)-(c) Intensity maps of (a) $E_{2g}$ mode, (b) $A_{1g}$ mode, and (c) PL. (d)-(f) The maps of (d) $E_{2g}$ frequency, (e) $A_{1g}$ frequency, and (f) PL energy. (g) PL and Raman spectra of two locations P1 and P2. (h) Schematic illustration of PL intensity inhomogeneity. (i) The FWHM map of PL.

would be 0.13%, 0.7%, and 0.17%, respectively, for $E_{2g}$, $A_{1g}$, and PL. Apparently, the estimated strain difference from $A_{1g}$ is much larger in magnitude than those from $E_{2g}$ and PL, suggesting that the doping effect, which preferentially affects $A_{1g}$, could play a significant role in the monolayer $MoS_2$ film. Notably, for $A_{1g}$ along the large wrinkle, it is blue-shifted with respect to the general area, in contrast to the red-shift for $E_{2g}$. Usually, the inhomogeneity of strain distribution can cause exciton localization as observed in both bulk and 2D materials, where PL is enhanced at lower band gap areas.[12, 18, 35] Such anti-correlation between the emission intensity and energy occurs under the conditions that the excitons are sufficiently mobile and non-radiative recombination rates are comparable between the high and low energy regions. On the contrary, the PL mapping data show generally positive correlation between the intensity (Figure 2c) and the peak energy (Figure 2d). For instance, the region P1 (P2) marked on Figure 2c-d, and f exhibits higher (lower) PL intensity and energy as well as $E_{2g}$ Raman frequency, as shown in Figure 2g. As pointed out above, the higher PL energy and Raman frequency is mainly attributed to the presence of more compressive strain or less tensile strain. The strain variation could be caused by the non-planar film morphology, which likely occurred in the film transfer process, in the forms of wrinkles and ripples as shown schematically in Figure 2h. The charge transfer between the film and substrate has been found to be sensitive to the details of the film – substrate contact.[1] At the interface of $SiO_2$ and $MoS_2$, interfacial states are formed due to the presence of high density dangling bonds on the surface of $SiO_2$. Electrons tend to be trapped in these states.[36, 37] In the non-rippled regions, the photo-generated non-equilibrium carriers will more likely be depleted through the film-substrate interface. In the rippled regions, where the PL energy is higher, in principle, the photo-generated carriers can drift to the regions with lower energies. However, because of the limited carrier mobility of the film and/or high carrier depletion rate in the lower energy regions, the PL intensity in the rippled regions turns out to be higher. Besides, the relative blue-shift of $A_{1g}$ mode along the large wrinkle, though it is not the focus of this work, could be also due to the less close contact of the film with the substrate, hence less n-type doping. Note that the PL peak position in the mapping (Figure 2f) is associated with $A^-$ trion rather than A exciton, because maximum PL peak position over the mapped area is 1.855 eV, implying that the $A^-$ emission dominates the PL emission. Figure 2i shows the map of PL full width at half maximum (FWHM) over the mapped area. The FWHM of the $A^-$ peak was found unaffected by the substrate;[21] and we also find a small variation of $115 \pm 5$ meV over the mapped area. These results indicate that the $MoS_2$ film has been n-type doped, and the origin of the doping could be the polymer residues left behind the transfer process.[14, 21]

Next, single point Raman measurements were carried out on one fixed location in the film (near the center of the mapped area) over a temperature range from RT to 305 °C with a step of 20 °C. Figure 3a shows temperature dependent Raman spectra, showing red-shifts of both $E_{2g}$ and $A_{1g}$ modes with increasing temperature. The change in peak position of the $E_{2g}$ and $A_{1g}$ modes with increasing temperature are plotted in Figure 3b-c, respectively. The $E_{2g}$ mode shows a very linear temperature dependence, which can be fitted well by a linear dependence:

$$\omega = \omega_0 + \chi \Delta T,$$

where $\omega_0$ is the mode frequency at RT, $\Delta T$ is the temperature change relative to RT, and $\chi$ is the first-order temperature coefficient.

However, the temperature dependence of $A_{1g}$ mode is rather nonlinear and can only be described by a third-order polynomial function used in our previous paper:[2]

$$\omega(T) = \omega_0 + \chi_1 \Delta T + \chi_2 (\Delta T)^2 + \chi_3 (\Delta T)^3,$$

where $\chi_1$, $\chi_2$, and $\chi_3$ are the first-, second- and third-order temperature coefficients. The temperature coefficients of the $E_{2g}$ and $A_{1g}$ modes obtained from the first temperature cycle are listed in Table 1, compared with the previously obtained bulk values. The results are in a good agreement with our previous work.[2] As discussed there, the nonlinear temperature dependence of $A_{1g}$ mode is attributed to the change in film morphology, while $E_{2g}$ mode is not sensitive to morphology. With increasing temperature, the film tends to change its morphology due to the mismatch in thermal expansion coefficients (TECs) between the $SiO_2$ substrate and the $MoS_2$ monolayer. At RT, $SiO_2$ TEC is ~$0.5 \times 10^{-6}$ K$^{-1}$, much smaller than that of monolayer $MoS_2$ ~$7 \times 10^{-6}$ K$^{-1}$, and increases with increasing temperature at a much smaller rate than $MoS_2$.[38, 39]

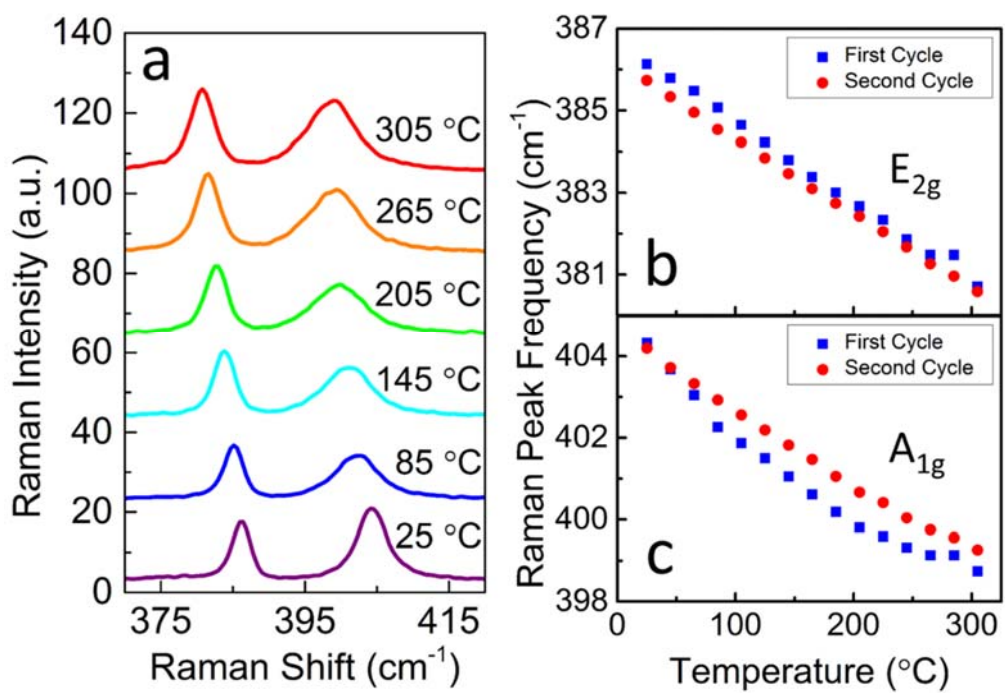

**Figure 3.** Raman data taken during two thermal annealing cycles. (a) Representative Raman spectra at selected temperatures in the first cycle. (b)-(c) Temperature dependence of (b) $E_{2g}$ and (c) $A_{1g}$ frequency for both cycles.

Therefore, the strain in MoS$_2$ film tends to accumulate with increasing temperature, and eventually goes beyond the confinement force of van der Waals bonding between the film and substrate, resulting in the change of the film morphology. Similar result has also been observed in graphene and WS$_2$.[4, 30] The film morphology change with increasing temperature led to the change in the mechanical coupling of MoS$_2$ film with SiO$_2$ substrate.[2, 4] With increasing temperature, the contact between the MoS$_2$ film and the substrate became more uniform and closer, leading to a greater extent of charge transfer between the film and the substrate or through interfacial states. The accelerated $A_{1g}$ red-shift with increasing temperature suggests increasing the equilibrium electron density, which could be due to enhanced charge injection from the substrate into the film and decomposition of adsorbed contaminants. Therefore, the first-cycle annealing process actually modified the film morphology and got rid of at least most polymer residues from the transfer process as well as adsorbed $H_2O$ and $O_2$.

**Table 1.** Temperature coefficients of $E_{2g}$ and $A_{1g}$ modes for both cycles.

|  | $E_{2g}$ | $A_{1g}$ | | |
|---|---|---|---|---|
|  | $\chi$ | $\chi^1$ | $\chi^2$ | $\chi^3$ |
| First cycle | $-0.0192 \pm 3.2 \times 10^{-4}$ | $-0.0390 \pm 3.5 \times 10^{-3}$ | $6.97 \times 10^{-5} \pm 2.4 \times 10^{-5}$ | $-3.69 \times 10^{-8} \pm 4.8 \times 10^{-8}$ |
| Second cycle | $-0.0183 \pm 8.6 \times 10^{-5}$ | $-0.0198 \pm 1.4 \times 10^{-3}$ | $-5.74 \times 10^{-6} \pm 9.8 \times 10^{-6}$ | $4.18 \times 10^{-8} \pm 2.0 \times 10^{-8}$ |
| Bulk (Ref. 2) | $-0.0221 \pm 8.9 \times 10^{-4}$ | $-0.0197 \pm 8.9 \times 10^{-4}$ | | |

After reaching 305 °C, the sample was annealed for 30 minutes and then cooled down to RT. The Raman and PL mappings were carried out again with the results shown in Figure 4. Except for along the long wrinkle, the intensity maps of both $E_{2g}$ and $A_{1g}$ modes (Figure 4a,b) become more uniform compared with those before the annealing (Figure 2a,b), while the overall PL intensity (Figure 4c) increases. As mentioned above, the annealing process can not only change the film morphology and strain but also burn off the polymer residues, accounting for the overall red-shift of $E_{2g}$ (Figure 4d) and blue-shift of PL energy (Figure 4f) after the annealing. The TEC mismatch between the film and the $SiO_2$ substrate will introduce tensile strain after cooling down to RT, leading to the red-shift of $E_{2g}$, although less significant for $A_{1g}$. However, the overall red-shift is actually more significant for $A_{1g}$, which should be explained as caused by the doping effect of broken-down polymer residues. As for PL, the removal of polymer residues and other contaminants on the film eliminates the non-radiative recombination channels to the excited carriers, leading to not only the blue-shift of PL energy but also the increase of PL intensity. The pattern of PL intensity map now becomes very different from that before the annealing (Figure 2c) and does not correspond to the pattern of $E_{2g}$ frequency map (Figure 4d) anymore, indicating that the doping effect has gradually turned into the dominant factor of the PL behavior. In fact, improved overall uniformity in the Raman frequency maps for both $E_{2g}$ and $A_{1g}$ seems to suggest that the small ripples that initially contributed to the PL intensity inhomogeneity have been mostly removed after the annealing.

We then performed the second-cycle temperature dependent Raman measurements at the same location as in the first cycle. The Raman frequency shifts of $E_{2g}$ and $A_{1g}$ modes are also plotted in Figure 3b-c to make a direct comparison with the first cycle. Similar to the results of the first cycle, the $E_{2g}$ mode shows a nearly linear temperature dependence, although with a slightly smaller slope than that in the first cycle, which can be explained by the small tensile strain created by the first cycle; while the $A_{1g}$ still remains nonlinear but to a much less extent than that in the first cycle, and its first-order temperature coefficient is almost halved compared to the first-cycle result, becoming very close to the bulk value. The fitting results are listed in Table 1. After the first-cycle annealing, the morphology of the film has been modified, i.e. the contact with the substrate has been improved, resulting in the changes in not only the doping concentration but also the strain distribution in the film. It is worth noting that annealing at 305 °C did not introduce damage or decomposition to $MoS_2$ monolayer, because the Raman intensity after two thermal cycles did not show a significant change, as shown in Figure S1 of Supporting Information.

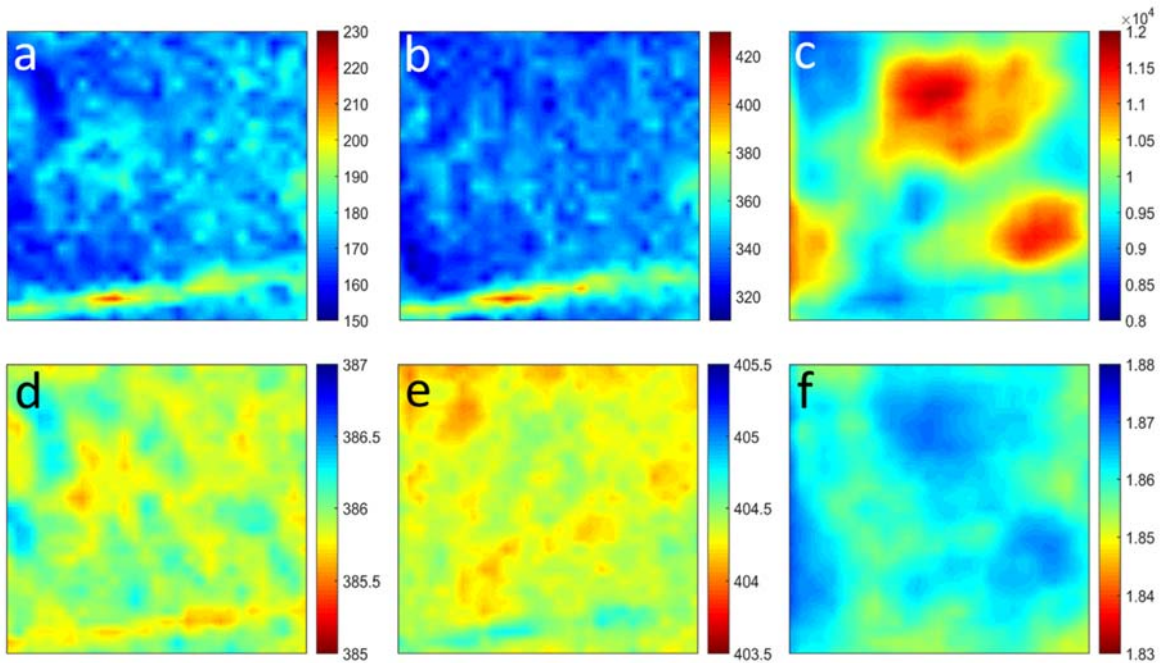

**Figure 4.** Room temperature Raman and PL mapping results after the first thermal annealing. (a)-(c) Intensity maps of (a) $E_{2g}$ mode, (b) $A_{1g}$ mode, and (c) PL. (d)-(f) The maps of (d) $E_{2g}$ frequency, (e) $A_{1g}$ frequency, and (f) PL energy.

At the end of the second round, the film was annealed at 305 °C for one hour to further remove the possible remaining polymer residues, then returned to RT. Raman and PL mappings were performed again at RT, and the results are shown in Figure 5 with mapping data of Raman and PL: peak positions (Figure 5a-c) and PL intensity (Figure 5d). The maximum spatial variations of Raman and PL peak positions are found to be ~1 cm$^{-1}$ for $E_{2g}$, ~0.7 cm$^{-1}$ for $A_{1g}$, and ~15 meV for PL, respectively, over the $MoS_2$ film. The $E_{2g}$ variation can be explained by the strain, yielding a range of ~0.22%, which is larger than that before the first cycle where it was due to the morphology fluctuation in the film and also that after the first cycle the bonding with the substrate has been created. By comparing to the mapping data before and after the second cycle, the $E_{2g}$ frequency map (Figure 4d vs. Figure 5a) on average exhibits a red-shift in Raman frequency, with the top part of mapped area showing more shift than the lower part. The difference could reflect the variation of film-substrate bonding strength. For the top part, the bonding is stronger so that annealing generates more tensile strain in the film when cooled down to RT. The $A_{1g}$ frequency map (Figure 5b) shows an overall blue-shift compared to that before the second cycle (Figure 4e), and becomes somewhat similar to that before the first annealing (Figure 2e), which could be explained by the charge exchange with the substrate and indicate that the charge exchange is rather sensitive to the film-substrate bonding. For the PL mapping, compared with the results after the first annealing cycle, the PL energy is overall further blue-shifted and the PL quantum yield is also further increased. After the second thermal annealing cycle, the remaining polymer residues on the surface were further removed, which in turn reduced the non-radiative recombination in the film, giving rise to overall higher PL quantum yields than those before the second cycle. Although improved contact with the substrate might enhance the carrier depletion through the substrate, the contaminants removal seems to be dominant in affecting the overall PL intensity. Figure 5e compares the PL spectra from locations L1 and L2 (marked in Figure 5c,d). The PL intensity at L2 is higher than that at L1 where the PL peak energy is lower. The peak energy of L2 is 1.87 eV, close to the A excitonic emission, indicating that the removal of polymer residues reduces the trion concentration in the film, hence an increase in the A excitonic recombination rate. Whereas, at L1, though the polymer residues have been removed after annealing, the closer contact of the film with the substrate would deplete the electrons through the interfacial states, resulting in much lower PL intensity than that at L2. Figure 5f and 5g compare the PL spectra of the two locations before and after the first annealing cycle and after the second annealing cycle. The comparison clearly shows that the PL peak energy gradually shifts from A⁻ transition to A after two thermal annealing cycles, and the PL intensity at L1 after the second cycle does not show a significant increase as that at L2 which almost quadruples the intensity. Figure 5h shows the FWHM map of PL after the second cycle, and the regions with higher (lower) PL peak energies exhibit smaller (larger) FWHMs. Compared to the FWHM map before the first cycle (Figure 2j), the FWHM at L1 decreases to 80 meV while that at L2 remains similar, which from another point of view supports the variation of doping concentration over the film. Improvement in free exciton PL characteristics (intensity and linewidth) typically reflects suppression of competing recombination channels and structural fluctuations, which usually leads to improved electronic conductivity. A correlative study on the optical and electrical response will be carried out in the future.

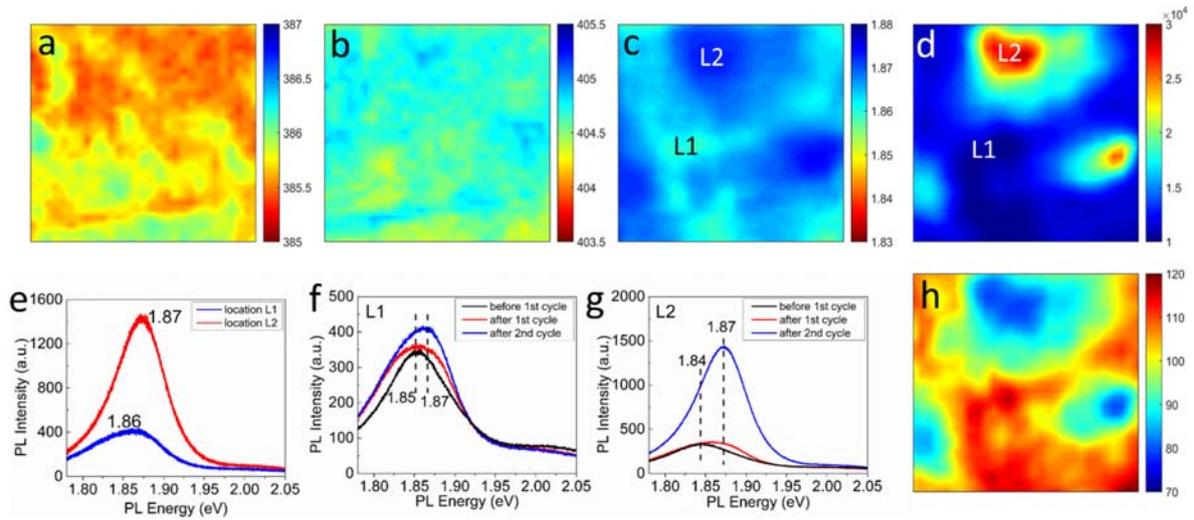

**Figure 5.** Room temperature Raman and PL mapping results after the second annealing cycle. (a)-(c) The maps of (a) $E_{2g}$ frequency, (b) $A_{1g}$ frequency, and (c) PL energy. (d) The map of PL intensity. (e) PL spectra of two locations L1 and L2. (f)-(g) The PL spectrum comparison with those before the first annealing cycle at (f) L1 and (g) L2. (h) The FWHM map of PL spectra.

An alternative way of showing the annealing effects is given in Supporting Information where Raman frequencies for $E_{2g}$ and $A_{1g}$, PL energy and intensity of the three mapping results, respectively shown in Figure 2, 4, and 5, are displayed together side-by-side for a more direct comparison (Figure S2).

**Conclusion**

In summary, we have carried out in-situ Raman probes in two thermal cycles as well as Raman and PL mapping before and after two-cycle annealing on a monolayer $MoS_2$ film transferred on a $SiO_2$/Si substrate to study the strain and doping effects on the electronic and optical properties of the monolayer $MoS_2$. Before annealing, the film-substrate bonding was weak and highly non-uniform along with the presence of chemical contaminants, where the inhomogeneous strain in the transferred film was the major cause of the fluctuations in Raman and PL peak position, and the maximum strain difference over the film was estimated to be ~0.13% by the phonon shift of $E_{2g}$ mode. However, after annealing the film-substrate bonding was significantly improved and the polymer residues were burned off, and the film-substrate bonding became the leading factor of the variations in Raman and PL peak positions and intensities. The strain inhomogeneity associated with the film-substrate bonding increased to ~0.22%. These findings suggest that annealing process can not only modify the film morphology and the film-substrate bonding, but also remove the polymer residues from the transfer process, and hence the optical and electronic performances of the $MoS_2$ films can be improved or altered.

## Experimental Methods

**Synthesis and Transfer of $MoS_2$.**

Monolayer $MoS_2$ was prepared using our previously reported CVD method with molybdenum chloride ($MoCl_5$) and sulfur as the precursor.[40] The $MoCl_5$ powder was placed at the center the furnace and sulfur at the upstream entry of the furnace, while the receiving substrates were placed downstream in a distance of 1-7 centimeters away from the center of the furnace. The furnace was heated up to at a rate of 28 °C/min with Ar gas purged. High quality and large area $MoS_2$ monolayer film was synthesized on sapphire wafer if proper parameters including temperature, Ar flow rate, and the amount of precursor were achieved.

The method used to transfer as-grown on-sapphire $MoS_2$ to a Si wafer coated with 300 nm $SiO_2$ was reported in our previous work with the assistance of PS.[31] A thin layer of PS was spin coated onto the as-grown sample, followed by a baking at 80-90 °C for 15 min to facilitate intimate adhesion of the PS layer with the $MoS_2$ film. With the assistance of a water droplet that penetrates all the way through the MoS$_2$ film, the PS-MoS$_2$ assembly was delaminated and transferred onto the SiO$_2$/Si substrate. After baking the transferred PS-MoS$_2$ assembly at proper temperature to remove the water residues, the PS was removed by rinsing with toluene several times.

**PL and Raman Measurements.**

μ-Raman and PL measurements were performed with a Horiba LabRAM HR800 system using a 532 nm excitation laser with a 50× long-working-distance lens (NA = 0.5), and the laser power used was ≤ 1 mW, sufficiently low not to cause significant shifting in both Raman modes. All the PL and Raman measurements were carried out in a Linkam TS1500 heating system. In the temperature dependent Raman measurement, N$_2$ gas was purged through the heating chamber at a very low flow rate to avoid the oxidation of MoS$_2$ film.[2,4] The temperature was elevated gradually to 305 °C with a step of 20 °C at a rate of 10 °C/min. At each temperature, the spectrum was acquired after allowing at least five minutes for thermal stabilization of the sample.

**Acknowledgements**

Y.Z. acknowledges the support of Bissell Distinguished Professorship. L.C. acknowledges the support of a Young Investigator Award from the Army Research Office (W911NF-13-1-0201).

Supporting Information

# In-Situ Monitoring of Thermal Annealing Induced Evolution in Film Morphology and Film-Substrate Bonding in a Monolayer MoS$_2$ Film

Liqin Su[1], Yifei Yu[2], Linyou Cao[2], and Yong Zhang[1,*]

[1]Department of Electrical and Computer Engineering, University of North Carolina at Charlotte, Charlotte, NC 28223

[2]Department of Materials Science and Engineering, North Carolina State University, Raleigh, NC 27695

* To whom correspondence should be addressed: yong.zhang@uncc.edu

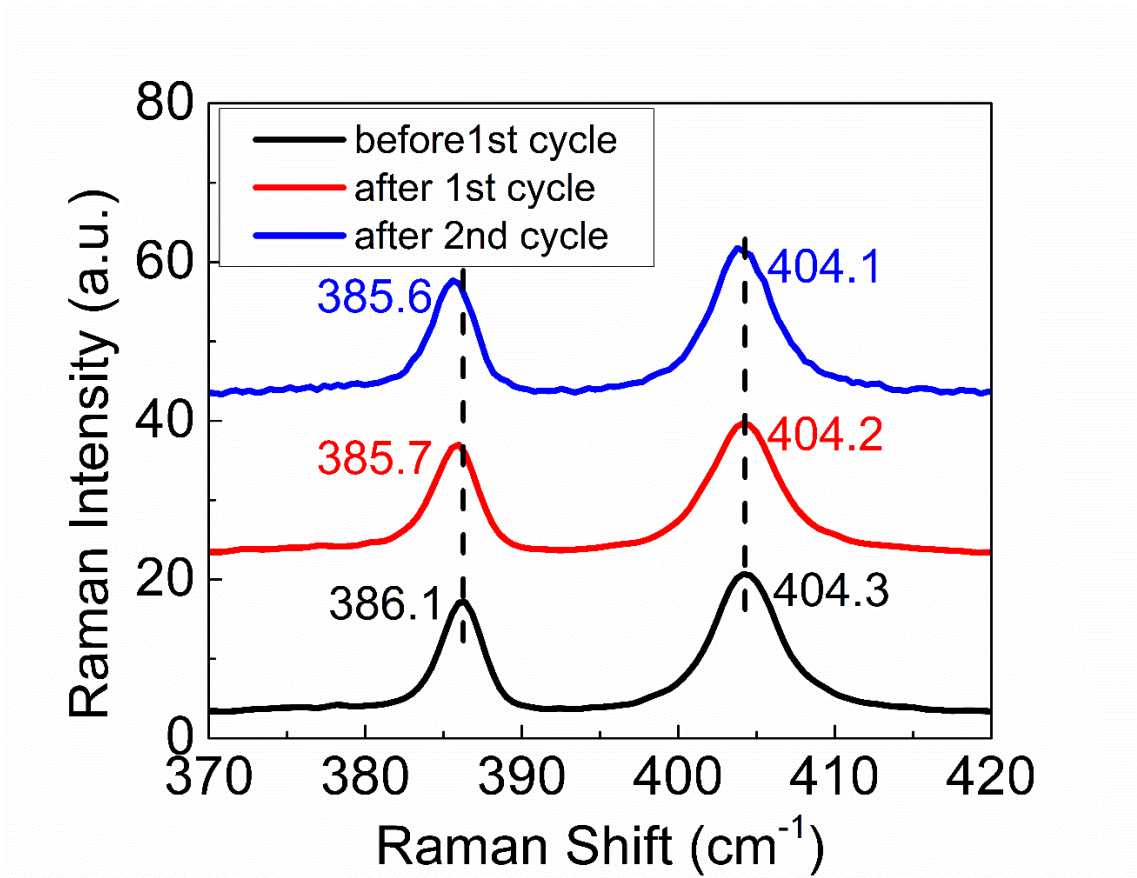

**Figure S1.** Comparison of Raman spectra taken before and after the first cycle and after the second cycle on the same location. The spectra were shifted vertically for clarity.

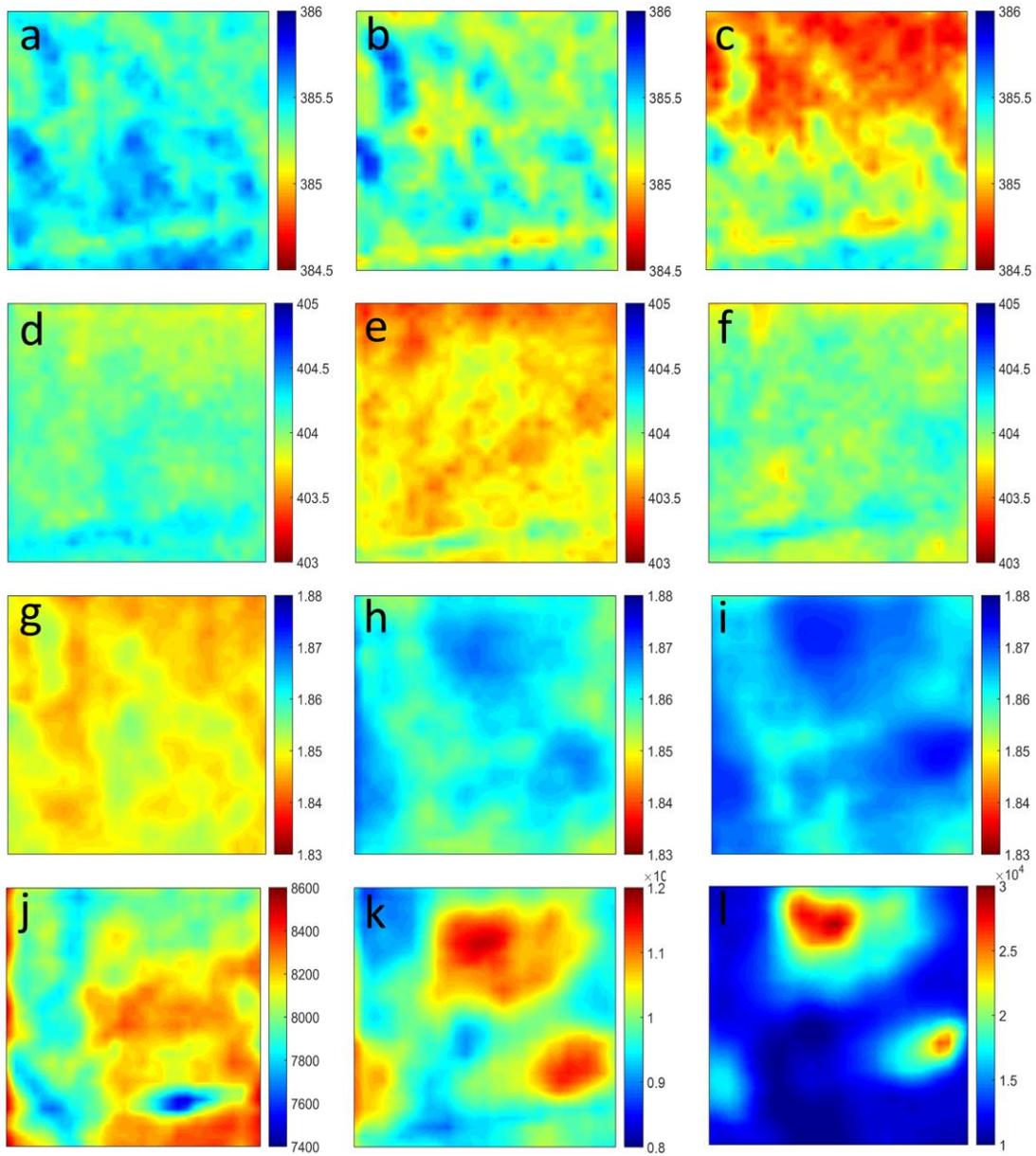

**Figure S2.** Side-by-side comparisons of Raman and PL mapping data before the first cycle (left column), before the second cycle (middle column), and after the second cycle (right column). (a)-(c) Maps of $E_{2g}$ Raman frequency. (d)-(f) Maps of $A_{1g}$ Raman frequency. (g)-(i) Maps of PL peak energy. (j)-(l) Maps of integrated PL intensity.